%% file: 00_GPsOverDensities.tex
\documentclass[letterpaper,conference,10pt]{IEEEtran}

\IEEEoverridecommandlockouts
\usepackage{bm,bbm} 
\usepackage{amsmath}
\usepackage{amsfonts}
\usepackage{amssymb}
\usepackage{dsfont} 
\usepackage{graphicx}
\usepackage[hidelinks]{hyperref}
\usepackage{url} 
\usepackage{color}

\usepackage{booktabs}

\usepackage{theorem}
\usepackage{subcaption}

\usepackage{paralist} 

\usepackage{flushend}
\usepackage[protrusion=false]{microtype}

\usepackage{multicol} 

\usepackage{etoolbox}
\let\bbordermatrix\bordermatrix
\patchcmd{\bbordermatrix}{8.75}{4.75}{}{}
\patchcmd{\bbordermatrix}{\left(}{\left[}{}{}
\patchcmd{\bbordermatrix}{\right)}{\right]}{}{}

\newcommand{\rv}[1]{\ensuremath{\boldsymbol{#1}}}
\newcommand{\nvec}[1]{\ensuremath{\underline{#1}}}

\newcommand{\mat}[1]{{\ensuremath{\mathbf{#1}}}}

\DeclareMathOperator{\xlog}{xlog}

\newcommand{\tr}{^{\top}}
\newcommand{\inv}{^{-1}}

\newcommand{\expect}[1]{{\rm E}\left\lbrace #1 \right\rbrace}

\newcommand{\IN}{\mathbb{N}}  
\newcommand{\IR}{\mathbb{R}}  

\newcommand{\klammer}[1]{\left( #1 \right)}

\ifx\d
\newcommand{\d}{{\rm\ d}}
\else
\renewcommand{\d}{{\rm\ d}}
\fi

\newcommand{\params}{\nvec{w}}
\newcommand{\dimx}{n}

\newcommand{\outval}[1]{y_{#1}}
\newcommand{\X}{\mat{X}}
\newcommand{\Y}{\nvec{y}}
\newcommand{\invec}[1]{\nvec{x}_{#1}}
\newcommand{\noise}[1]{\rv{\nu}_{#1}}
\newcommand{\D}{\mathfrak{D}}
\newcommand{\inpdf}[1]{\mathfrak{d}_{#1}}

\newcommand{\GP}{\mathcal{GP}}
\newcommand{\Kernel}{\mat{\Sigma}}

\newcommand{\xsys}{\nvec{x}}
\newcommand{\ysys}{\nvec{y}}

\title{\Large \bf A Distance-based Framework for Gaussian Processes over Probability Distributions}

\author{%
	\IEEEauthorblockN{{\bf Maxim~Dolgov}}
	\IEEEauthorblockA{
		Robert Bosch GmbH\\
		Corporate Research\\
		{maxim.dolgov@bosch.com}
	}\and
	\IEEEauthorblockN{{\bf Uwe~D.~Hanebeck}}
	\IEEEauthorblockA{
		Intelligent Sensor-Actuator-Systems Laboratory (ISAS) \\
		Institute for Anthropomatics and Robotics \\ 
		Karlsruhe Institute of Technology (KIT), Germany \\ 
		{uwe.hanebeck@ieee.org}
	}
}

\begin{document}
	\maketitle
	\thispagestyle{empty}
	\pagestyle{empty}
	
	
	\begin{abstract}
		Gaussian processes constitute a very powerful and well-understood method for non-parametric regression and classification. In the classical framework, the training data consists of deterministic vector-valued inputs and the corresponding (noisy) measurements whose joint distribution is assumed to be Gaussian. In many practical applications, however, the inputs are either noisy, i.e., each input is a vector-valued sample from an unknown probability distribution, or the probability distributions are the inputs. In this paper, we address Gaussian process regression with inputs given in form of probability distributions and propose a framework that is based on distances between such inputs. To this end, we review different admissible distance measures and provide a numerical example that demonstrates our framework.
	\end{abstract}

	\allowdisplaybreaks[1]
	
	\section{Introduction}
	\label{sec:Introduction}
\input{Sections/01_Introduction}
	\section{Proposed Framework}
	\label{sec:GPs}
	\input{Sections/02_GPs}
	
	\section{Distance Measures for Probability Distributions}
	\label{sec:Distances}
\input{Sections/02_Distances}
		
	\section{Numerical Example}
	\label{sec:NumericalExample}
\input{Sections/03_NumericalExample}
	\section{Conclusion}
	\label{sec:Conclusion}
	\input{Sections/04_Conclusion}

	

	
	
	

%
	%
	
	
	\bibliographystyle{IEEEtranNoURL}
	\bibliography{Sections/00_Literature}
	
\end{document}

%% file: Sections/01_Introduction.tex
The problem of learning a regression or classification function given a training dataset can be addressed by either a \emph{parametric} or a \emph{nonparametric} approach. In the parametric approach, the designer selects a function model, e.g., a linear function or a neural network, and optimizes a single set of parameters of the model such that the model fits the training data. In many cases, choosing the model before the data is available leads to poor performance. A natural approach in this case would be to increase the number of function parameters. However, this step bears the risk of overfitting, where the performance on the training data becomes very good but the results for the test data are very poor. In nonparametric learning, the designer still has to choose a model family. But instead of having a single fixed set of parameters, the number of parameters either grows with the data or there are infinitely many parameter sets with an assigned probability distribution~\cite{Murphy_2012}. The methods from the latter class are referred to as Bayesian.
\begin{figure}[t]
	\centering
	\includegraphics[width=.4\textwidth]{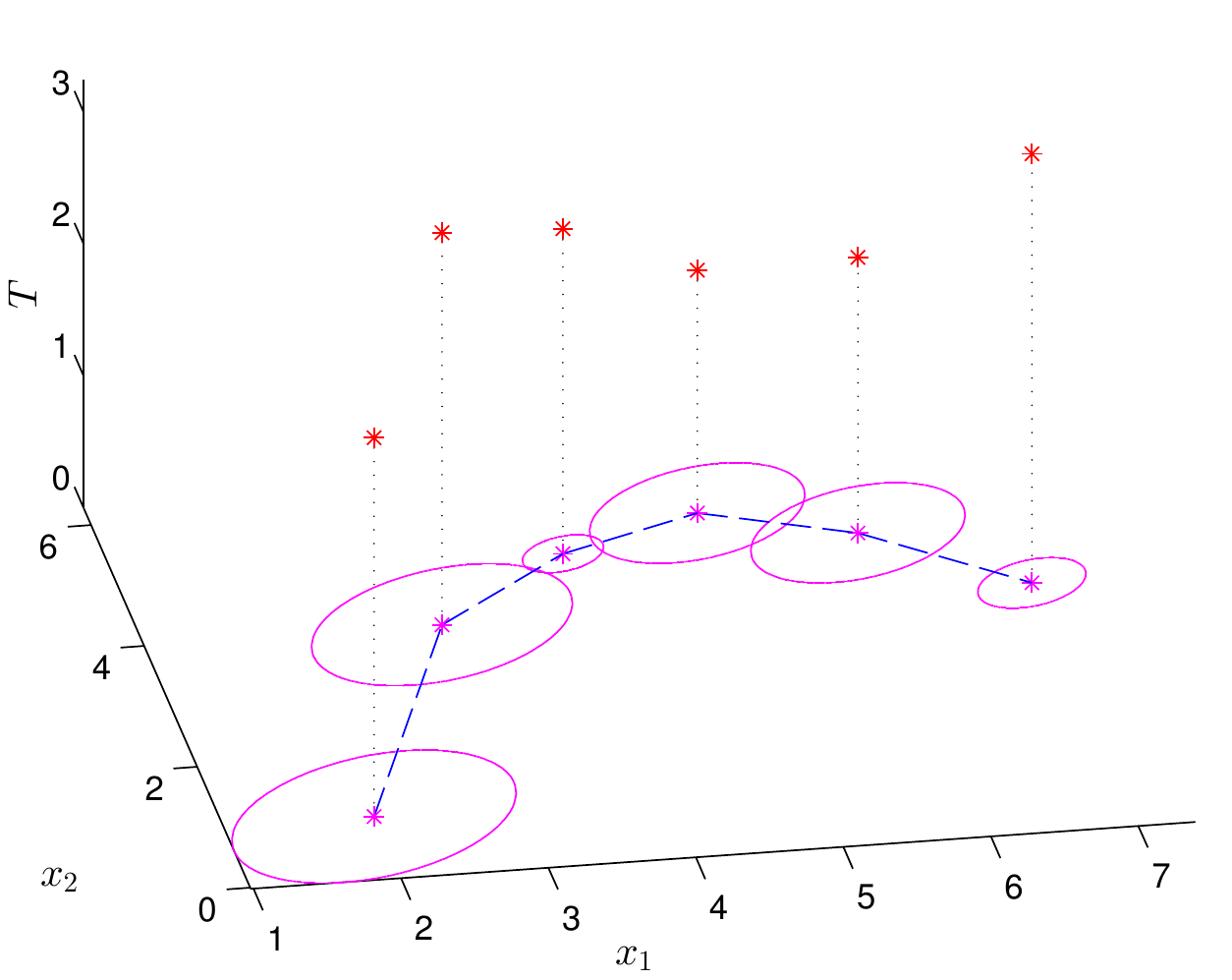}
	\caption{Illustration of noisy temperature measurements (red stars) collected with a mobile robot. The true robot path is represented by the blue dashed line, while the magenta stars and ellipses depict the mean and the covariance of the estimated pdf of the robot position in the plane.}
	\label{fig:SensorScenario}
\end{figure}

Gaussian Processes (GPs) constitute a special class of Bayesian nonparametric methods where the probability distribution of the model parameters, and the probability distribution of the measurements given the inputs and the parameters are Gaussian~\cite{Rasmussen_2005}. Moreover, the mean and the covariance of the Gaussian of the outputs are functions of the inputs. However, the classical GP formulation does not provide a consistent way to address problems where the inputs are noisy, i.e., vector-valued samples drawn from unknown underlying probability distributions, or the inputs are the probability distributions themselves. 
Consider for example the following scenarios. We wish to learn a model of a temperature distribution on a plane based on noisy temperature measurements collected using a mobile robot. In the first scenario, the robot can measure its position, however, these position measurements are corrupted by noise. Thus, the inputs to the GP that is used to learn the temperature distribution are noisy vectors. Now, assume that the robot uses the position measurements or measurements from, e.g., an inertial measurement unit or landmark measurements in order to estimate its position. The position estimates given in form of probability distributions then serve as the inputs to the GP\footnote{An alternative formulation of the problem could be to use the entire sequences of measurements from which the position estimates are inferred as the inputs to the GP. In this case, however, the GP needs to be able to deal with such input data, which is even more challenging than having probability distributions as inputs because the dimension of the inputs is not constant.}. This scenario is depicted in Fig.~\ref{fig:SensorScenario}.

The distinction between the two described input classes (noisy inputs and probability distributions) is very important. In the case of noisy inputs, the input to the GP is still a vector although its value does not correspond to the value of the true input that generated the measurement. In the second case, where the input to the GP is a probability distribution, the input is a function, i.e., an infinite-dimensional quantity, and therefore not admissible to classical GP framework. Of course, one could argue that in case of probability distributions provided as inputs it is possible to use, e.g., the means as the inputs GP. However, by doing so, we lose information about the uncertainty. Furthermore, the mean may not be the appropriate input representation, e.g., if the probability distribution is multi-modal.


In this paper, we address GPs whose inputs are probability distributions and propose a framework that is based on distances between probability distributions. The remainder of the paper is organized as follows. First, we give a brief introduction to the classical GP framework with deterministic inputs, discuss approaches to GPs with probability distributions as inputs, and summarize the contribution of the paper. In Sec.~\ref{sec:GPs}, we present a new framework for GPs that are defined over distances between probability distributions and review different admissible distance measures in Sec.~\ref{sec:Distances}. A numerical example demonstrates the proposed framework in Sec.~\ref{sec:NumericalExample} and Sec.~\ref{sec:Conclusion} concludes the paper.

%

\subsection{Gaussian Processes with Deterministic Inputs}
\label{subsec:GPsDeterministicInputs}

In what follows, we give a brief introduction to GPs with deterministic inputs. For a much more thorough introduction, the interested reader is referred to~\cite{Rasmussen_2005}. 

Consider the measurement model
\begin{align}
	y
		&=
		a(\invec{}) + \noise{}\ ,
\end{align}
where $y\in\IR$ denotes the measurement, $\invec{}\in\IR^\dimx$ the input vector, and $\noise{}$ is an independent and identically distributed Gaussian noise with zero mean and variance~$\sigma_\nu^2$. The measurement function $a(\cdot)$ is not known. However, we assume that given a finite set of inputs $\invec{i}$, $i=1,2,\ldots,N$, $N\in\IN$, the corresponding measurements $y_i$ and every subset thereof are jointly normally distributed. Then, the measurements $y_i$ are said to be generated by a GP with mean function $m(\invec{}) = \expect{a(\invec{})}$ and covariance function $k(\invec{i},\invec{j}) = \expect{(a(\invec{i})-m(\invec{}))(a(\invec{j})-m(\invec{}))\tr}$ that depends on the parameters $\params\in\IR^m$. We then write
\begin{align*}
	\begin{bmatrix}
		y_1 & \ldots & y_N
	\end{bmatrix}\tr \sim \GP_{\params}(\nvec{\mu},\mat{\Sigma})\ .
\end{align*}
with
\begin{align*}
	\nvec{\mu} 
		= 
		\begin{bmatrix} 
			m(\invec{1}) \\ m(\invec{2}) \\ \vdots \\ m(\invec{N})
		\end{bmatrix} ,\ 
	\mat{\Sigma} = 
		\begin{bmatrix}
			k(\invec{1},\invec{1}) & \hspace{-3mm}k(\invec{1},\invec{2}) & \hspace{-3mm}\ldots & \hspace{-3mm}k(\invec{1},\invec{N})\\
			k(\invec{2},\invec{1}) & \hspace{-3mm}k(\invec{2},\invec{2}) & \hspace{-3mm}\ldots & \hspace{-3mm}k(\invec{2},\invec{N})\\
			\vdots & \hspace{-3mm}\vdots & \hspace{-3mm}\ddots & \hspace{-3mm}\vdots\\
			k(\invec{N},\invec{1}) & \hspace{-3mm}k(\invec{N},\invec{2}) & \hspace{-3mm}\ldots & \hspace{-3mm}k(\invec{N},\invec{N})
		\end{bmatrix} ,
\end{align*}
where $\Kernel$ is referred to as the kernel of the GP.

Given a training set $(\X,\Y)$,  where $\X=\{\invec{1},\ldots,\invec{N}\}$ and $\Y=\{\outval{1,\ldots,\outval{N}}\}$, the goal is to predict the output $\outval{*}$ at a test input $\invec{*}$. For the probability density function (pdf) of $\outval{*}$, the so-called predictive density of $\outval{*}$, it holds
\begin{align*}
	p(\outval{*}|\invec{*},\X,\Y)
		&=
		\int\limits_{\IR^m} p(\outval{*}|\invec{*},\params) p(\params|\X,\Y) \d\params\ ,
\end{align*}
where $p(\params|\X,\Y)$ is the posterior (Gaussian) pdf of the parameters $\params$ given the training data $(\X,\Y)$ and the parameter prior $p_0(\params)$. The parameters $\params$ of a GP are also referred to as the \emph{hyperparameters}. If we do not have the prior $p_0(\params)$, the hyperparameters $\params$ can be estimated by maximizing the likelihood $p(\Y|\X,\params)$ with respect to $\params$.
%

Since all involved pdfs are assumed to be Gaussian, the prediction of $\outval{*}$ at $\invec{*}$ is determined by the mean $\mu_*$ and covariance $\sigma^2_*$ that are given by
\begin{align}
\begin{aligned}
	\mu_*
		&=
		k(\invec{*},\X)\tr\Kernel\inv\Y\ ,\\
	\sigma^2_*
		&=
		k(\invec{*},\invec{*}) - k(\invec{*},\X)\tr\Kernel\inv k(\invec{*},\X)\ .
\end{aligned}
\label{eq:Prediction}
\end{align}
Please note that although \eqref{eq:Prediction} has a quite simple formulation, its evaluation can be computationally intense if the set of training data and/or the dimension of the inputs are large.

\subsection{Gaussian Processes over Probability Distributions}
\label{subsec:GPsNoisyInputs}

So far, the input $\invec{*}$ was assumed to be deterministic. As motivated above, many real-world applications of GPs and also the nonparametric (Bayesian) methods require to be able to deal with inputs provided in form of probability distributions. A general treatment of such inputs in Bayesian nonparametric methods is discussed in~\cite{Dellaportas_1995}.

In the context of GPs, the foundation for consideration of probability distributions as inputs was laid in~\cite{Girard_2002} where it was applied to prediction of time series. In this work, the authors considered the case where the training inputs $\invec{}$ were deterministic vectors and the test inputs were normal distributions. By doing so, the training of the GP can be performed by standard means, while the predictive density of the output $\outval{*}$ for an input provided in form of a probability distribution $\inpdf{*} = p_*(\invec{})\in\IR^\dimx$ can be computed according to
\begin{align}
	p(\outval{*}|\inpdf{*},\X,\Y)
		&=
		\int\limits_{\IR^n} p(\outval{*}|\nvec{x},\X,\Y) p_*(x) \d \nvec{x}\ ,
	\label{eq:PredictionGirard}
\end{align}
where $p(\outval{*}|\nvec{x},\X,\Y)$ is Gaussian as defined in Sec.~\ref{subsec:GPsDeterministicInputs}. The authors of~\cite{Girard_2002} further argue that an analytical evaluation of the integral in~\eqref{eq:PredictionGirard} is generally not tractable because $p(\outval{*}|\nvec{x},\params)$ is a complicated function of $\nvec{x}$. For this reason, they propose two approximation schemes. In the first scheme, only the mean and covariance of the predictive density are computed under the assumption that the probability distributions that are provided as inputs are Gaussian and the covariance function is the Squared Exponential (SE) covariance function~\cite{Rasmussen_2005}. For this case, an approximate expression for the mean and the variance of the predictive density is provided in~\cite{Girard_2002}. The exact solution can be found in~\cite{Candela_2003} (see also~\cite{Girard_2004,Candela_2004,Girard_2005}). The second approximation scheme consists in solving~\eqref{eq:PredictionGirard} using a Monte-Carlo approach according to
\begin{align}
p(\outval{*}|\inpdf{*},\X,\Y)
	&\approx
	\frac{1}{T} \sum\limits_{t=1}^{T} p(\outval{*}|\invec{t},\X,\Y)\ ,
\label{eq:PredictionMonteCarlo}
\end{align}
where $\nvec{x}_t$ are samples drawn from $\inpdf{*}$. In order to remain in the framework of GPs, i.e., in order to have Gaussian predictive densities, only the mean and covariance of~\eqref{eq:PredictionMonteCarlo} are maintained.

Another application of the first approximation method is considered in~\cite{Dallaire_2009}, where it is applied to stabilization of a nonlinear system. Here, not only the test inputs but also the training inputs are assumed to be Gaussian pdfs, i.e., the training data now becomes $(\D,\Y) = \{(\inpdf{1},\outval{1}),\ldots,(\inpdf{N},\outval{N})\}$ and the predictive density of $\outval{*}$ given the probability distribution $\inpdf{*} = p_*(\invec{})$ is given by
\begin{align}
p(\outval{*}|\inpdf{*},\D,\Y)
	&=
	\int\limits_{\IR^n} p(\outval{*}|\invec{},\D,\Y) p_*(\invec{}) \d \invec{}\ .
\label{eq:PredictionNoisy}
\end{align}
In~\cite{Dallaire_2009}, the problem of training the GP model and making predictions is addressed by introducing the mean covariance function
\begin{align}
	k(\inpdf{i},\inpdf{j})
		&=
		\int\limits_{\IR^n} \int\limits_{\IR^n} k(\invec{i},\invec{j}) p_i(\invec{i})p_j(\invec{j})\d\invec{i}\d\invec{j}\ .
	\label{eq:MeanKernel}
\end{align}
For this reason, we will refer to GPs constructed using the approach from~\cite{Dallaire_2009} as \emph{mean-kernel} GPs. As a special case, the authors of~\cite{Dallaire_2009} consider the SE covariance function and provide an analytical expression for~\eqref{eq:MeanKernel} (recall that the input distributions are assumed to be Gaussian). However, it is pointed out that estimating the hyperparameters of the GP using the maximum likelihood approach is not trivial because of the many local maxima of the log-likelihood~$\log p(\nvec{y}|\X,\params)$. To avoid this issue, the authors propose to apply the Maximum A Posteriori (MAP) approach by defining a (Gaussian) prior for the hyperparameters.

A different approach to training and predicting with Gaussian inputs was presented in~\cite{McHutchon_2011}. The authors propose to use a Taylor expansion of the measurement model
\begin{align*}
	\outval{}
		&=
		a(\widetilde{\invec{}} + \nvec{\nu}_x) + \nu_y
\end{align*}
in $\widetilde{\invec{}}$, where $\nvec{\nu}_x\in\IR^\dimx$ is the Gaussian input noise and $\nu_y\in\IR$ the Gaussian measurement noise. Although the derivatives in the expansion are again GPs~\cite{Solak_2003} and the first and the second derivatives can be calculated in closed form, an exact evaluation is computationally inefficient. For this reason, the authors propose to apply approximations. This allows to derive a linear model for the input noise. However, the training of the GP for this (approximate) linear model is still non-trivial and requires further approximations.

Finally, another method that allows to use GPs with Gaussian inputs and its application to predictions with localization uncertainty is considered in~\cite{Jadaliha_2013}. The authors address the problem by two approaches: the Monte-Carlo and the Laplace approximations. While the Monte-Carlo approximation is standard, the application of the Laplace approximation in the context of GPs is new. The Laplace approximation is used to compute an approximation of the integral in~\eqref{eq:PredictionNoisy}. However, the evaluation of the Laplace approximation~\cite{Tierney_1986} is computationally intense. Therefore, the authors propose further approximations.

\subsection{Contribution}

As we have seen, considering GPs with probability distributions as inputs is non-trivial and each existing approach has its disadvantages. First of all, the reviewed approaches consider only inputs that are Gaussian. But, this limitation can be avoided by using the Monte-Carlo approach and maintaining only the first and the second moments for the prediction, which implicitly approximates the output density with a Gaussian. Then, although the approach from~\cite{Girard_2002} is efficient, it does not allow for probability distributions as training inputs, while learning of the GP parameters of the method from~\cite{Dallaire_2009} is not simple even in the considered case of Gaussian distributions as training and test inputs. The approach from~\cite{McHutchon_2011} has to make approximations both in the measurement model and the learning of the GP parameters. Finally,~\cite{Jadaliha_2013} has also to make approximations in order to solve~\eqref{eq:PredictionNoisy} and an additional approximation to reduce the computational cost. For this reasons, we propose a novel framework for Gaussian Processes with inputs provided in form of probability distributions that defines the covariance function directly in the space of (arbitrary) probability distributions. Furthermore, our approach is not limited to a specific family of probability distributions such as Gaussians. Moreover, it even allows to incorporate both continuous and Dirac probability distributions, i.e., discrete distributions over a continuous domain, into the same GP. Finally, the Gaussian property of the predictive density is implicitly ensured by construction and does not involve approximations.

%% file: Sections/02_GPs.tex
In this section, we present the proposed framework. As outlined above, the main notion of our framework consists in defining the covariance function of the GP directly in the space of probability distributions. The proposed approach primarily applies to stationary covariance functions, i.e., covariance functions that are defined over the distances between the inputs. However, we also discuss how non-stationary covariance functions can be implemented within the proposed framework.

\subsection{Stationary Covariance Functions}
\label{subsec:StationaryCovFunc}
\begin{table}[t]
	\centering
	\begin{tabular}{ll}
		\toprule
		{\bf covariance function}& {\bf expression} \\
		\midrule
		constant & $\sigma_0^2$\vspace{1mm}\\
		squared exponential\hspace{.5cm} & $\exp\klammer{-\frac{1}{2}\frac{\Delta^2}{l^2}}$\vspace{1mm}\\
		Mat\'{e}rn & $\frac{2^{1-\nu}}{\Gamma(\nu)}\klammer{\frac{\sqrt{2\nu}}{l}\Delta}^\nu K_\nu\klammer{\frac{\sqrt{2\nu}}{l}\Delta}$\vspace{1mm}\\
		exponential & $\exp\klammer{-\frac{\Delta}{l}}$\vspace{1mm}\\
		$\gamma$-exponential & $\exp\klammer{-\klammer{\frac{\Delta}{l}}^\gamma}$\vspace{1mm}\\
		rational quadratic & $\klammer{1+\frac{\Delta^2}{2\alpha l^2}}^{-\alpha}$\vspace{1mm}\\
		\bottomrule
	\end{tabular}			
	\vspace{2mm}
	\caption{Covariance functions defined in terms of the distance $\Delta=d(\inpdf{i},\inpdf{j})$ between the probability distributions $\inpdf{i}$ and $\inpdf{j}$.}
	\label{tab:CovFuncTable}
\end{table}

To address the issue of GP inputs provided in form of probability distributions, we propose to \emph{use stationary covariance functions that take the distance between two probability distributions as the argument}. As a motivation for this approach, consider for example the classical isotropic SE covariance function
\begin{align*}
	k(\invec{i},\invec{j})
		&=
		\alpha^2\exp\klammer{-\frac{1}{2}\frac{(\invec{i}-\invec{j})\tr(\invec{i}-\invec{j})}{l^2}}
\end{align*}
defined for deterministic inputs $\invec{i}$ and $\invec{j}$, where ${\alpha,l\in\IR}$ are the hyperparameters. Here, the quadratic term ${(\invec{i}-\invec{j})\tr(\invec{i}-\invec{j})}$ corresponds to the squared Euclidean distance between $\invec{i}$ and $\invec{j}$. Now, instead of the vector-valued inputs $\invec{i}$ and $\invec{j}$, let the inputs be probability distributions $\inpdf{i}$ and $\inpdf{j}$. Then, we can redefine the SE covariance function according to
\begin{align*}
	k(\inpdf{i},\inpdf{j})
	&=
	\alpha^2\exp\klammer{-\frac{1}{2}\frac{d(\inpdf{i},\inpdf{j})^2}{l^2}}\ ,
\end{align*}
where $d(\inpdf{i},\inpdf{j})$ denotes any admissible distance measure between the input densities $\inpdf{i}$ and $\inpdf{j}$. The proposed design approach can be applied to any stationary covariance function. Table~\ref{tab:CovFuncTable} provides a small overview of selected stationary covariance functions from~\cite{Rasmussen_2005} redefined in terms of the distance between probability distributions. A discussion of a set of selected admissible distance measures between probability distributions is given given in Sec.~\ref{sec:Distances}.

Please note that the choice of stationary covariance functions defined over distances of probability distribution as the main basis of the proposed framework is justifiable because the space of probability distributions is not ordered or partially ordered. Therefore, non-stationary covariance function with absolute positions in the underlying space may not be necessary. Nevertheless, we discuss a method how such covariance functions still can be incorporated in the proposed framework in Sec.~\ref{subsec:NonStationaryCovFunc}.

It is worth noting that our approach somewhat resembles the approach from~\cite{Dallaire_2009} that uses the mean covariance function~\eqref{eq:MeanKernel}. However, in the method from~\cite{Dallaire_2009} it is necessary to compute the integrals in~\eqref{eq:MeanKernel} in each iteration step of the optimization problem that is solved in order to determine the hyperparameters. In the proposed approach however, the computation of the distances between the input densities is independent of the hyperparameters and only has to be performed once. Moreover, the estimation of the hyperparameters remains the same as in the classical GP framework.

\subsection{Non-stationary Covariance Functions}
\label{subsec:NonStationaryCovFunc}

A direct usage of non-stationary covariance functions is not possible within the presented framework. Hence, we propose the following workaround. According to~\cite{Rasmussen_2005}, it is possible to construct new covariance functions from existing ones using operations such as addition, multiplication, convolution, tensor product, etc. For construction of GPs with probability distributions provided as inputs and non-stationary covariance functions, we thus propose to combine stationary covariance functions with non-stationary functions that operate, e.g., on the means or the modes of the probability distributions provided as inputs. For example, the covariance function constructed from a linear function and the SE function according to
\begin{align*}
	k(\inpdf{i},\inpdf{j})
		&=
		\expect{\invec{i}}\tr\Sigma_d\expect{\invec{j}} \exp\klammer{-\frac{1}{2}\frac{d(\inpdf{i},\inpdf{j})^2}{l^2}}
\end{align*}
is admissible, where $\Sigma_d$ is a matrix of hyperparameters and the expectations are computed with respect to $\inpdf{i}$ and $\inpdf{j}$. More possible candidates for construction of non-stationary covariance functions can be found in~\cite{Rasmussen_2005}.

%% file: Sections/02_Distances.tex
In this section, we provide a small, not necessarily complete overview of distance measures between multivariate probability distributions that can be used in the proposed framework. In particular, we analyze the following distances
\begin{compactitem}
	\item total variation and $L_P$ distance,
	\item Hellinger distance,
	\item Jensen--Shannon divergence,
	\item Wasserstein/OSPA distance, and
	\item modified Cram\'{e}r--von Mises distance.
\end{compactitem}
A more thorough review of distance measures for probability distributions can be found, e.g., in~\cite{Zolotarev_1983}. In what follows, we not only present the distances listed above but also point out the classes of probability distributions that can be compared using the presented distances, i.e., whether the measures can be used to compute the distance between two continuous distributions, two Dirac mixture distributions, or a continuous and a Dirac mixture probability distribution. 
A Dirac mixture or particle distribution with $m$ components is given by
\begin{align*}
	f(\xsys)
		&=
		\sum\limits_{i=1}^{m} w_i \delta(\xsys-\xsys_i)\ ,
\end{align*}
where $0<w_i\leq 1$ are the weights with $\sum_{i=1}^{m} w_i=1$ and $\xsys_i$ are the positions of the Dirac components. Such densities are very important, e.g., in robotics and nonlinear filtering.
%
\subsubsection{Total Variation and $L_p$ Distance}\hfill

The total variation distance~\cite{Zolotarev_1983} of two continuous probability distributions $f$ and $g$ with respect to a measure $\mu$ is defined according to
\begin{align*}
	d_p(f,g)
		&=
		\klammer{\int_{\Omega} (f(\xsys)-g(\xsys))^p\ \text{d}\mu(\xsys)}^\frac{1}{p}\ .
\end{align*}
If we set $\text{d}\mu(\xsys)=\text{d}\xsys$, we obtain the $L_p$ distance between $f$ and $g$. Both these distance measures are defined only for two continuous distributions.
%
\subsubsection{Hellinger Distance}\hfill

The Hellinger distance between two continuous probability distributions $f$ and $g$ is given by
\begin{align*}
	d(f,g)
		&=
		\klammer{1-\int \sqrt{f(\xsys)g(\xsys)}\ \text{d}\xsys}^2\ .
\end{align*}
For this distance, it holds $0\leq d(f,g)\leq 1$. Therefore, it is less suitable for application in the proposed framework compared to other distances that are unbounded. Furthermore, there is no counterpart of the Hellinger distance for two Dirac distributions or a continuous and a Dirac distribution.

\subsubsection{Jensen--Shannon Divergence}\hfill

The Jensen--Shannon divergence that was introduced in~\cite{Lin_2006} is a symmetric version of the Kullback--Leibler divergence~\cite{Kullback_1951}. For the continuous probability distributions $f$ and $g$, it can be computed according to
\begin{align*}
	d(f,g)
		&=
		\frac{1}{2} \int f(\xsys)\log\frac{f(\xsys)}{p(\xsys)} + g(\xsys)\log\frac{g(\xsys)}{p(\xsys)}\ \text{d}\xsys\ ,
\end{align*}
where $p(\xsys) = (f(\xsys)+g(\xsys))/2$. Comparison of two Dirac distributions or a continuous and a Dirac distributions is not possible with this measure.
%
\subsubsection{Wasserstein/OSPA Distance}\hfill

For the Wasserstein distance~\cite{Wasserstein_1969} of two continuous probability distributions $f$ and $g$, it holds
\begin{align*}
	d_p(f,g)
		&=
		\inf_h \klammer{\int d_e(\xsys,\ysys)^p h(\xsys,\ysys)\ \text{d}\xsys\ \text{d}\ysys}^\frac{1}{p}\ ,
\end{align*}
where $d_e(\xsys,\ysys)$ is the Euclidean distance between the vectors $\xsys$ and $\ysys$, and $h(\xsys,\ysys)$ is a joint distribution whose marginals are $f(\xsys)$ and $g(\ysys)$, i.e., it holds $\int h(\xsys,\ysys)\ \text{d}y=f(\xsys)$ and $\int h(\xsys,\ysys)\ \text{d}\xsys = g(\ysys)$.

The analog of the Wasserstein distance between continuous distributions for two Dirac distributions with equal numbers of components, also referred to as the Optimal MAss Transfer~(OMAT) metric~\cite{Schuhmacher_2008}, is presented in~\cite{Hoffman_2002}. For the special case of $m$ equally weighted Diracs, it holds
\begin{align*}
	d_p(f,g)
		&=
		\klammer{\frac{1}{m}\min_{\pi\in\Pi} \sum\limits_{i=1}^{m} d(\xsys_i,\ysys_{\pi_i})^p}^\frac{1}{p}\ ,
\end{align*}
where $\Pi$ is the set of all possible assignments between the Diracs from the two distributions. An extension of the Wasserstein distance to Dirac distributions with different numbers of components is presented in~\cite{Schuhmacher_2008}, where it is referred to as the Optimal Sub-Pattern Assignment (OSPA) metric.

Please note that the Wasserstein metric is generally intractable for continuous distributions, because the infimum has to be taken over all possible joint distributions~$h$. Furthermore, the OMAT distance may not be efficiently computable for Dirac distributions with large numbers of components because it requires the solution of a linear assignment problem.

\subsubsection{Modified Cram\'{e}r--von Mises Distance}\hfill

The distance metrics for probability distributions presented so far are either limited to the same probability distribution class and cannot be used to compute the distance between a continuous and a Dirac distribution, or are intractable (Wasserstein distance for continuous distributions). For this reason, we propose to use the modified Cram\'{e}r--von Mises distance (mCvMd)~\cite{Hanebeck_2008,Hanebeck_2009}.

In order to present the mCvMd between two arbitrary $n$-dimensional probability distributions $f$ and $g$, we first introduce the notion of the Localized Cumulative Distribution~$F(\nvec{m},b)$ of $f$ according to
\begin{align*}
F(\nvec{m},b)
	&=
	\int\limits_{\IR^n} f(\xsys) k(\xsys,\nvec{m},b)\ \text{d}\xsys\ ,
\end{align*}
where $k(\xsys,\nvec{m},b)$ is a Radial Basis Function (RBF)
\begin{align*}
	k(\xsys,\nvec{m},b)
		&=
		\exp\klammer{-\frac{1}{2}\frac{(\xsys-\nvec{m})\tr(\xsys-\nvec{m})}{b^2}}\ .
\end{align*}
Then, the mCvMd of $f$ and $g$ can be calculated according to
\begin{align*}
	d(f,g)
		&=
		\klammer{\int\limits_{0}^{b_{\max}} \int\limits_{\IR^n} w(b)(F(\nvec{m},b)-G(\nvec{m},b))^2\ \text{d}\nvec{m}\ \text{d}b}^\frac{1}{2}
\end{align*} 
where $G(\nvec{m},b)$ is the LCD of $g(\xsys)$ and $w(b)$ is a weighting function with
\begin{align*}
	w(b)
		=
		\begin{cases}
			\frac{1}{b^{n-1}} & \text{for }b\leq b_{\max}\ ,\\
			0 & \text{otherwise}\ ,
		\end{cases}
\end{align*}
and $b_{\max}$ is a large positive constant.

According to~\cite{Hanebeck_2015}, the mCvMd for two Dirac distributions $f$ and $g$ with samples at $\xsys^f_i$ and $\xsys^g_j$, weights $w^f_i$ and $w^g_j$, $i=1,\ldots,M$, $j=1,\ldots,L$, and a large $b_{\max}$ evaluates to
\begin{align*}
	d(f,g)^2
		&=
		\frac{\pi^{\frac{n}{2}}}{8}\klammer{D_f-2D_{fg}+D_g + 2c_bD_E}\ ,
\end{align*}
with $c_b=\log\klammer{4b_{\max}^2}-\Gamma$ and
\begin{align*}
	D_f
		&=
		\sum\limits_{i=1}^{M}\sum\limits_{j=1}^{M} w^f_iw^f_j \xlog\klammer{(\xsys^f_i-\xsys^f_j)\tr(\xsys^f_i-\xsys^f_j)}\ ,\\
	D_{fg}
		&=
		\sum\limits_{i=1}^{M}\sum\limits_{j=1}^{L} w^f_iw^g_j \xlog\klammer{(\xsys^f_i-\xsys^g_j)\tr(\xsys^f_i-\xsys^g_j)}\ ,\\
	D_g
		&=
		\sum\limits_{i=1}^{L}\sum\limits_{j=1}^{L} w^g_iw^g_j \xlog\klammer{(\xsys^g_i-\xsys^g_j)\tr(\xsys^g_i-\xsys^g_j)}\ ,\\
	D_E
		&=
		\klammer{\sum\limits_{i=1}^{M}w^f_i\xsys^f_i - \sum\limits_{j=1}^{L}w^g_j\xsys^g_j}\hspace{-3mm}{\vphantom{\klammer{\sum\limits_{i=1}^{M}w^f_i\xsys^f_i - \sum\limits_{j=1}^{L}w^g_j\xsys^g_j}}}\tr\hspace{-1mm} \klammer{\sum\limits_{i=1}^{M}w^f_i\xsys^f_i - \sum\limits_{j=1}^{L}w^g_j\xsys^g_j} ,
\end{align*}
where
\begin{align*}
	\xlog(x)
		&=
		\begin{cases}
			0&\text{for }x=0\ ,\\
			x\log(x)&\text{otherwise}\ .
		\end{cases}
\end{align*}
A more thorough discussion of the mCvMd can be found in~\cite{Gilitschenski_2015}.

%% file: Sections/03_NumericalExample.tex
\begin{figure}[t]
	\centering
	\begin{subfigure}[c]{.48\textwidth}
		\includegraphics[width=\textwidth]{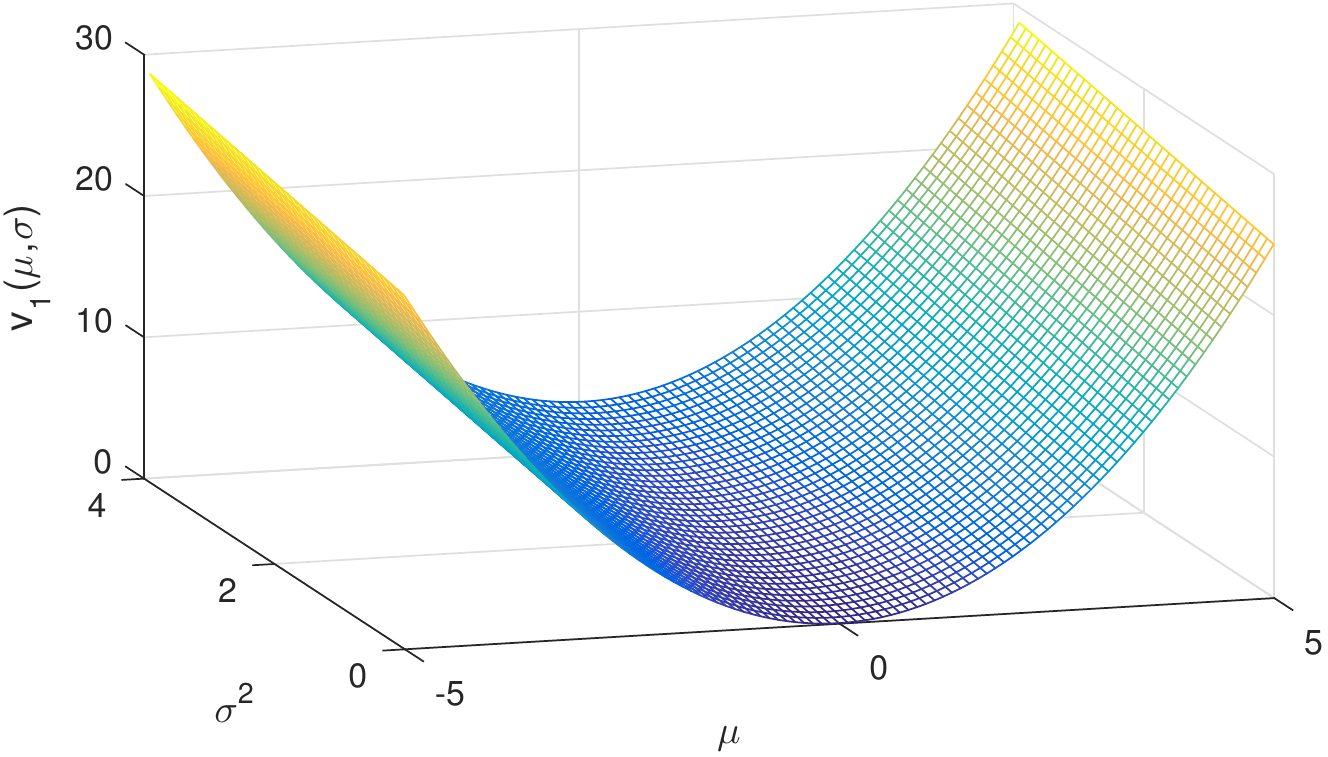}
		\subcaption{Quadratic function $v_1(\mu,\sigma)$}
	\end{subfigure}
	\begin{subfigure}[c]{.48\textwidth}
		\includegraphics[width=\textwidth]{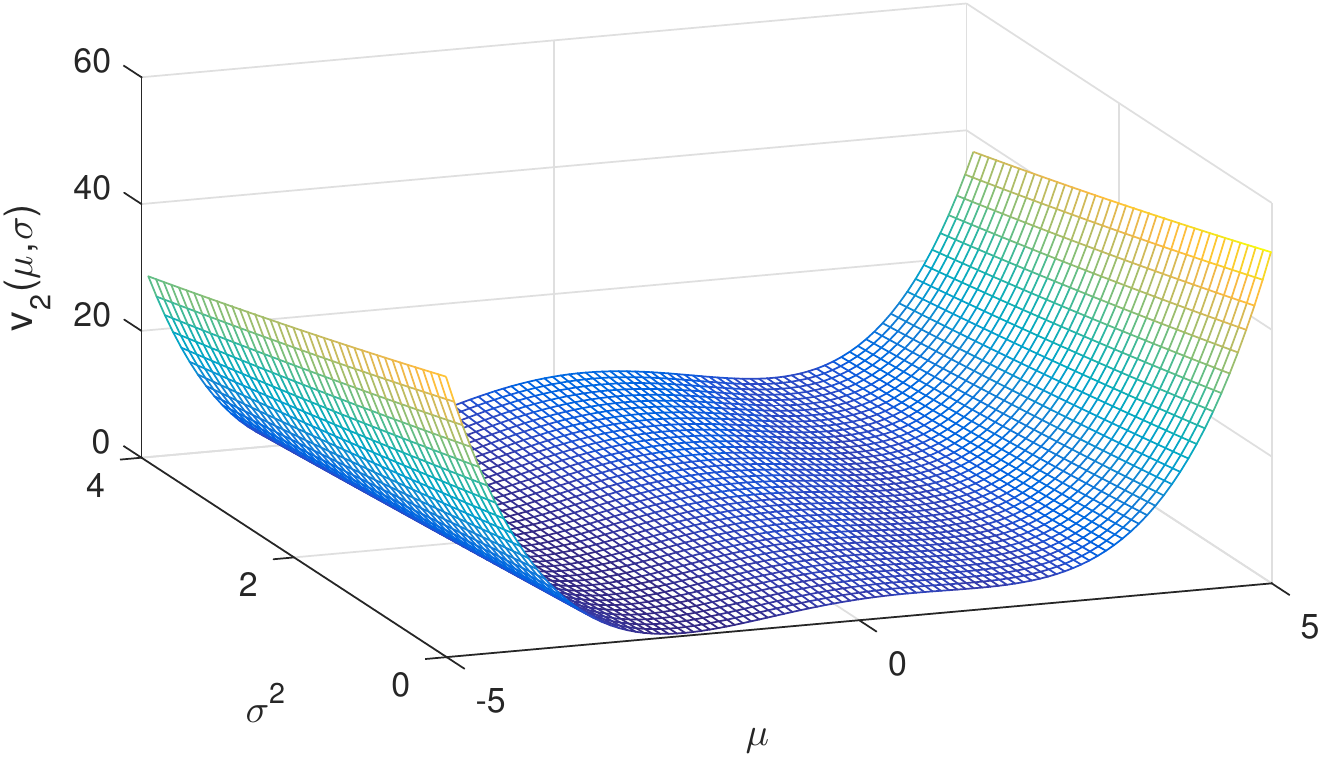}
		\subcaption{Rosenbrock function $v_2(\mu,\sigma)$}
	\end{subfigure}
	\caption{Functions $v_1(\mu,\sigma)$ and $v_2(\mu,\sigma)$ that are used in the numerical regression example.}
	\label{fig:Functions}
\end{figure}

\begin{align}
	\widetilde{v}_1(\xsys)
		&=
		\expect{\xsys\tr\xsys}\nonumber\\
		&=
		\expect{\xsys}\tr\expect{\xsys} + \expect{(\xsys-\expect{\xsys})\tr(\xsys-\expect{\xsys})}\nonumber\\
		&=
		\mu_x\tr\mu_x + \sigma_x^2\ ,
	\label{eq:Quadratic}
\end{align}
where $\mu_x$ is the mean of $\xsys$ and $\sigma_x^2$ its variance, and a slightly modified version of the Rosenbrock function\footnote{We added the parameter $d$.}
\begin{align}
\widetilde{v}_2(x_1,x_2)
	&=
	a(x_1+b)^2 + c(x_1^2-x_2+d)^2\ .
	\label{eq:Rosenbrock}
\end{align}
where we set $x_1=\mu_x$, $x_2=\sigma_x^2$, $a=c=0.1$, $b=4$, and $d=-4$.
%
For the sake of visualization, we perform this demonstration over univariate Gaussians with means in the range $\mu\in[-5, 5]$ and covariances $\sigma^2\in[0.1^2, 2^2]$. The two functions~\eqref{eq:Quadratic} and~\eqref{eq:Rosenbrock} then become
\begin{align*}
	v_1(\mu,\sigma)
		&=
		\mu^2+\sigma^2\ ,\\
	v_2(\mu,\sigma)
		&=
		0.1(\mu+4)^2 + 0.1(\mu^2-\sigma^2-4)^2\ .
\end{align*}
The functions $v_1(\mu,\sigma)$ and $v_2(\mu,\sigma)$ are depicted in Fig.~\ref{fig:Functions}.
In this section, we demonstrate the proposed framework applied to regression of two functions, a simple quadratic function

\begin{figure}[t]
	\centering
	\includegraphics[width=.4\textwidth]{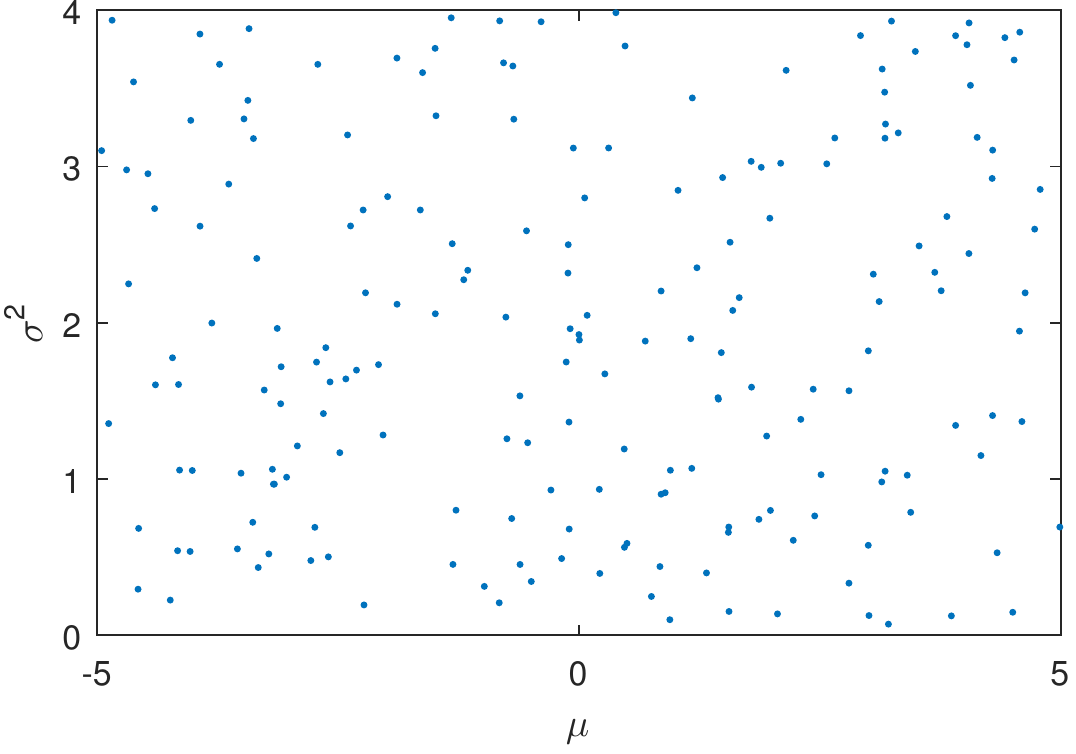}
	\caption{Distributions used as training inputs. Each point represents a training Gaussian with mean $\mu$ and variance $\sigma^2$.}
	\label{fig:trainGaussians}
\end{figure}

In particular, we proceed as follows. First, we draw $200$ Gaussians for training from the considered range by drawing a mean and a variance and compute the outputs for each distribution using $v_1(\mu,\sigma
)$ or $v_2(\mu,\sigma)$, respectively. The sampled Gaussians are depicted in Fig.~\ref{fig:trainGaussians}, where each point represents a Gaussian with its corresponding mean and variance. For the simulation, we chose to represent the inputs to the GP as Dirac mixture distribution in order to include the GP constructed using the proposed framework with the Wasserstein distance. For this reason, we draw samples from the training Gaussians that are then used as inputs to the GP, i.e., each input is a set of samples drawn from one of the Gaussians that are used for training. The $10$ samples from the Gaussians are drawn deterministically using the method from~\cite{Hanebeck_2009}. In summary, the training data for the GP consists of $200$ sets \`a $10$ samples and a function value for each sample set. We use this training data in order to train a GP that is based on the mean kernel~\cite{Dallaire_2009} (here we used the original sampled mean and variances of the Gaussians), and two GPs designed according to the proposed framework, where one GP uses the mCvMd and the other the Wasserstein distance.
\begin{figure*}[h]
	\centering
	\begin{subfigure}[c]{.31\textwidth}
		\includegraphics[width=\textwidth]{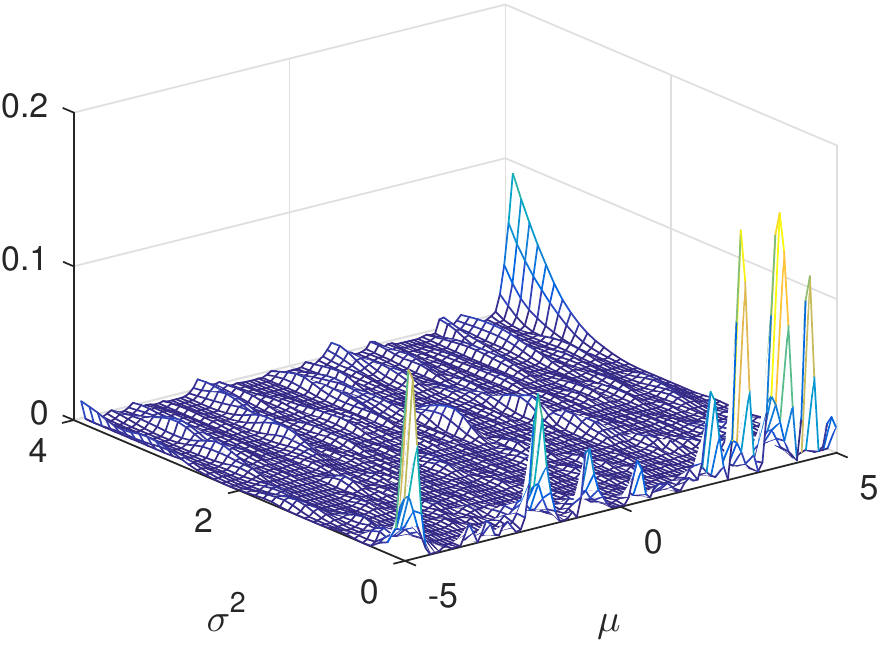}
		\subcaption{proposed -- mCvMd}
	\end{subfigure}
	\hspace{4mm}
	\begin{subfigure}[c]{.31\textwidth}
		\includegraphics[width=\textwidth]{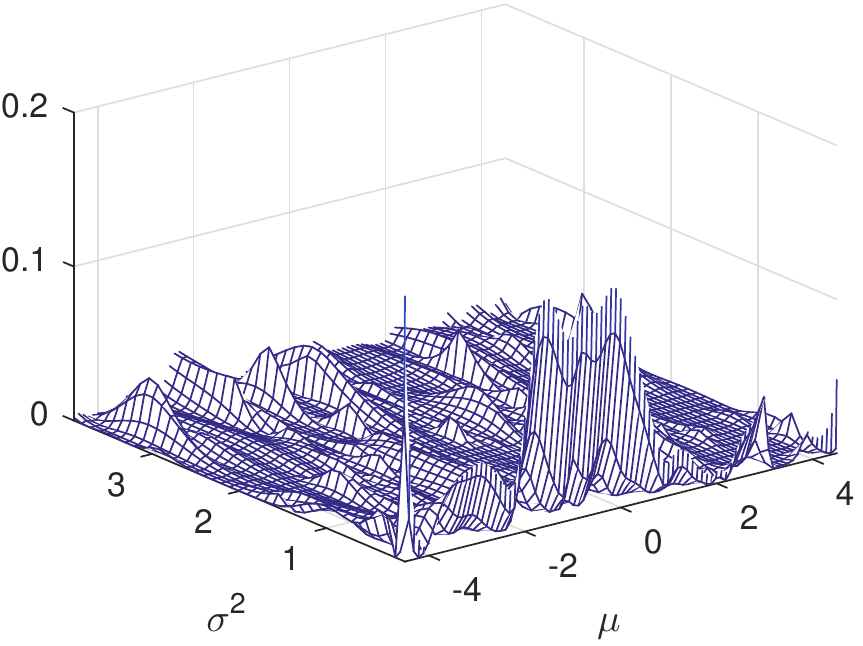}
	\subcaption{proposed -- Wasserstein distance {(cropped to ${\mu\in[-4.5, 4.5],}~{\sigma^2\in[.1, 3.9]}$)}}
	\end{subfigure}
	\hspace{4mm}
	\begin{subfigure}[c]{.31\textwidth}
		\includegraphics[width=\textwidth]{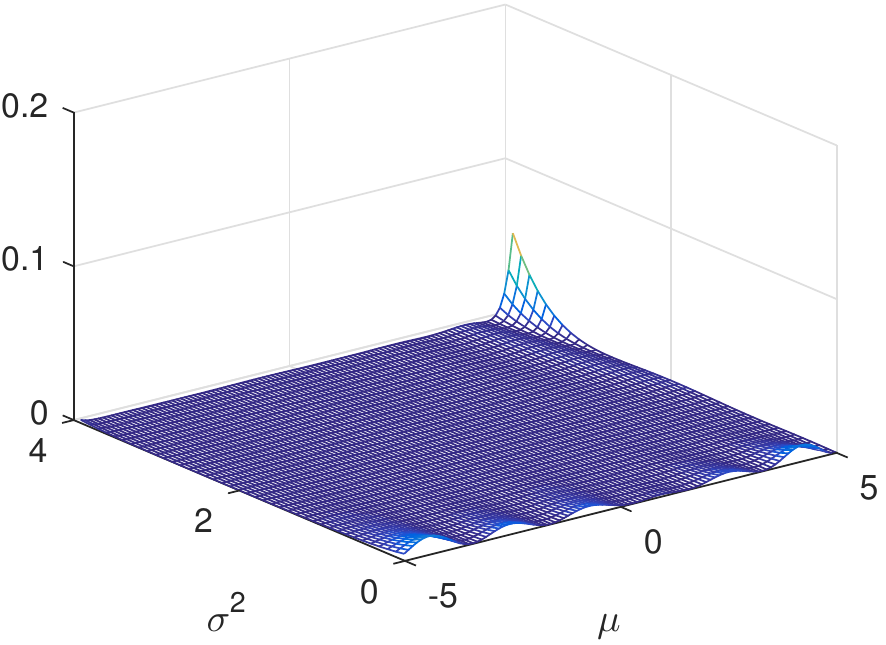}
		\subcaption{mean kernel~\cite{Dallaire_2009}}
	\end{subfigure}
	\caption{Quadratic error of the GP regression of the quadratic function $v_1(\mu,\sigma)$.}
	\label{fig:SimQuadratic}
\end{figure*}
\begin{figure*}[h]
	\centering
	\begin{subfigure}[c]{.31\textwidth}
		\includegraphics[width=\textwidth]{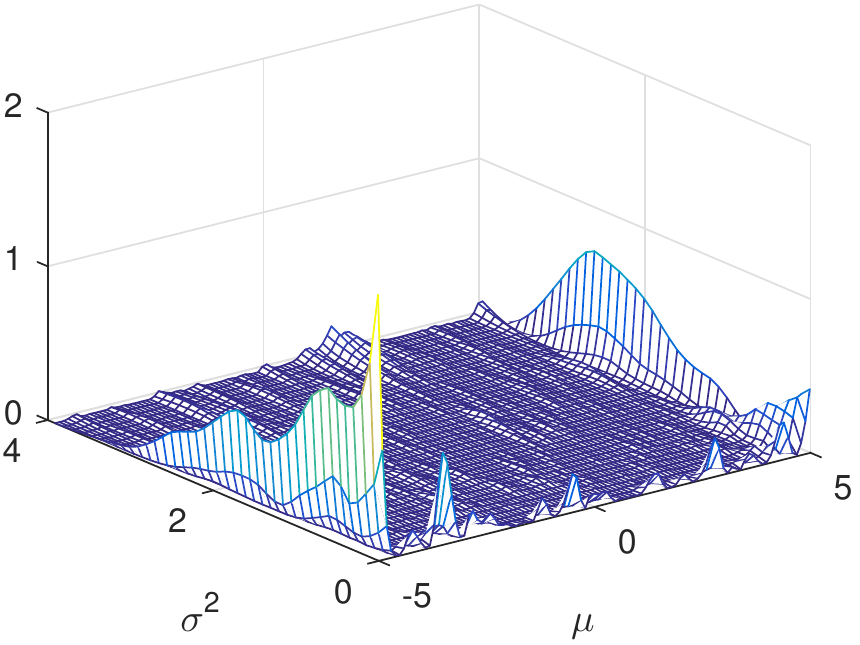}
		\subcaption{proposed -- mCvMd}
	\end{subfigure}
	\hspace{4mm}
	\begin{subfigure}[c]{.31\textwidth}
		\includegraphics[width=\textwidth]{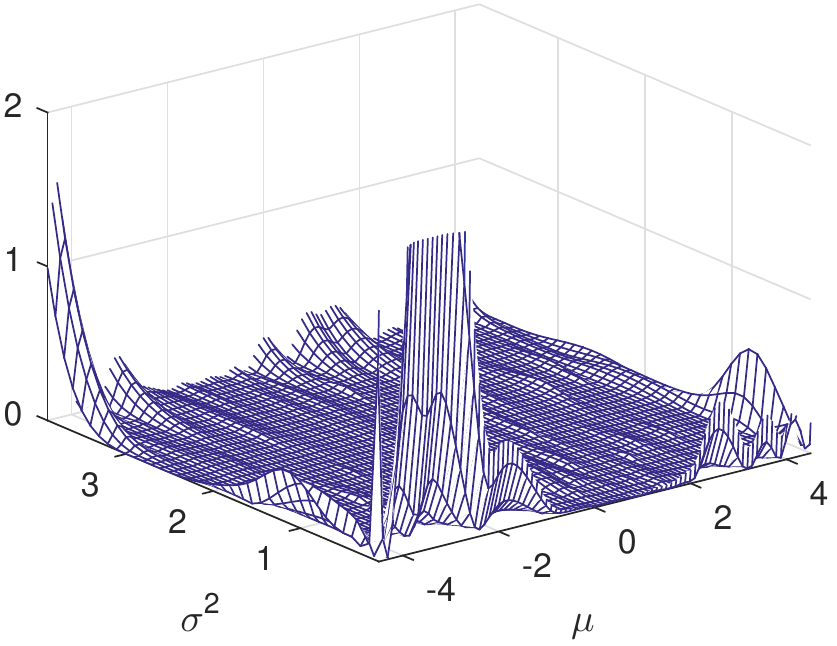}
		\subcaption{proposed -- Wasserstein distance {(cropped to ${\mu\in[-4.5, 4.5],}~{\sigma^2\in[.1, 3.9]}$)}}
	\end{subfigure}
	\hspace{4mm}
	\begin{subfigure}[c]{.31\textwidth}
		\includegraphics[width=\textwidth]{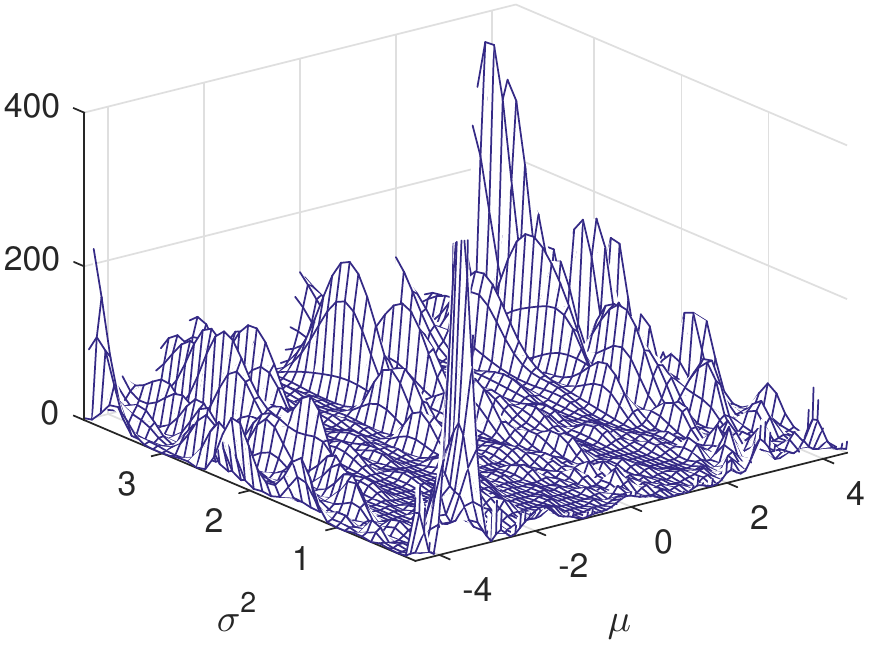}
		\subcaption{mean kernel~\cite{Dallaire_2009} {(cropped to ${\mu\in[-4.5, 4.5],}~{\sigma^2\in[.1, 3.9]}$)}}
	\end{subfigure}
	\caption{Quadratic error of the GP regression of the Rosenbrock function $v_2(\mu,\sigma)$.}
	\label{fig:SimRosenbrock}
\end{figure*}

The simulation results can be seen in Figs.~\ref{fig:SimQuadratic} and~\ref{fig:SimRosenbrock} that depict the quadratic errors of the regressions of the quadratic function $v_1(\mu,\sigma)$ and the Rosenbrock function $v_2(\mu,\sigma)$. Some of the figures have been cropped due to relatively large errors on the boundaries. In Fig.~\ref{fig:SimQuadratic}, it can be seen that the quadratic function $v_1(\mu,\sigma)$ can be approximated well by all three GPs, whereas the GP based on the mean kernel slightly outperforms the GP designed with the proposed framework that uses the mCvMd. The GP based on the Wasserstein distance performs worst but its performance is still comparable with the two other GPs. 

In the scenario where we analyze the approximation of the Rosenbrock function $v_2(\mu,\sigma)$ (Fig.~\ref{fig:SimRosenbrock}), the GP based on the mCvMd performs best and the GP based on the Wasserstein distance is only slightly worse. However, as in the regression of the quadratic function, the performance of the GP based on the Wasserstein distance is bad on the boundaries. For this reason, we suggest to use the mCvMd rather than the Wasserstein distance in the proposed framework. The mean-kernel GP~\cite{Dallaire_2009} performs much worse than the GPs designed using the proposed distance-based framework. This is probably due to the wrong estimation of the hyperparameters that was performed using the maximum likelihood approach. From this issue, we may conclude that our framework is much more convenient in practice because the estimation of the hyperparameters is the same as in the classical GP framework with deterministic vector-valued inputs and therefore no prior is required. Furthermore, the estimation of the hyperparameters for the GP constructed according to our framework is faster because the distances between probability distributions that are provided as inputs are independent of the hyperparameters. On the other hand, the integrals in~\eqref{eq:MeanKernel} have to be evaluated in each iteration step of the optimization algorithm that estimates the hyperparameters of the mean-kernel GP from~\cite{Dallaire_2009}.

A reference implementation of the proposed algorithm is available on GitHub~\cite{Dolgov_GPframework}.

%% file: Sections/04_Conclusion.tex
In this paper, we proposed a framework for GPs, where the inputs are provided in form of probability distributions. The main notion of the proposed framework is to use stationary covariance functions that take the distances between the input probability distributions as arguments. We further discussed how it is possible to construct GPs with non-stationary kernels and compared several admissible distance measures. The proposed framework has the advantage that it can operate with arbitrary probability distributions if the appropriate distance measure is chosen. Moreover, we are able to construct GPs whose inputs contain continuous probability distributions and probability distributions represented using particles. In our numerical example, we compared our approach with an existing state-of-the-art method that is based on the notion of the mean kernel. The approximation quality of the GPs designed according to our framework was good for the considered quadratic and the Rosenbrock functions. Especially the GP based on the mCvMd performed well. In regression of the Rosenbrock function, our framework outperformed the mean-kernel approach from~\cite{Dallaire_2009}.